Short Paper*

# First principles calculations of the electronic properties of O- and $O_2$-$NbSe_2$ complexes

Alvanh Alem G. Pido
Department of Physics, Mindanao State University – Marawi Campus, Philippines
alvanhalem.pido@msumain.edu.ph
(corresponding author)

Bryan P. Pagcaliwagan
Department of Physics, Mindanao State University – Marawi Campus, Philippines
pagcaliwagan.bp46@s.msumain.edu.ph

## Abstract

*Purpose* – We investigated the interaction of O and $O_2$ on monolayer Niobium Diselenide ($NbSe_2$) to provide theoretical predictions about the electronic properties of the complexes using First principles calculations in Quantum Espresso 6.7. As known, considering impurities in pristine nanomaterials like $NbSe_2$ is very important as it can alter some of its properties.

*Method* – In this paper, we performed some topological analyses on the electronic densities and electronic structures calculations to O- and $O_2$-$NbSe_2$ complexes. Charge Density Difference (CDD) and Bader charge analysis reveal that O and $O_2$ acted as oxidizing agents and accumulated electronic charges from the $NbSe_2$.

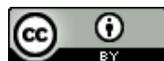

*Results* – The electronic properties calculations of the complexes showed that the metallic behavior of $NbSe_2$ is preserved after O and $O_2$ adsorption. Calculations of the net charge transfer revealed that the atomic and molecular oxygen has accumulated electronic charges while the $NbSe_2$ has depleted electronic charges. These results showed the possibility of tailoring the electronic properties of $NbSe_2$.

*Conclusion* – The interaction of O and $O_2$ with the monolayer $NbSe_2$ caused charge redistributions while maintaining the metallicity of the $NbSe_2$. In all circumstances, the results are consistent with the established works which show the possibility of modifying the electronic properties of $NbSe_2$ that could open some potential applications in nanotechnology and other nanoelectronics-related devices.

*Recommendation* – Further calculations could be done like electron localization function, optical properties, and vibrational properties, to understand more the nature of their interaction.

*Practical Implication* – This study provided insights and theoretical predictions on the electronic properties of oxygen adsorption to $NbSe_2$ which could help to better understand how metallic nanomaterials like $NbSe_2$ react with oxygen, leading to some potential applications.

## INTRODUCTION

Niobium diselenide ($NbSe_2$) is one of the emerging 2D materials nowadays which is naturally metallic at room temperature (Xi et al., 2015; Yeoh et al., 2020), and shows a charge density formation in the monolayer (Nguyen et al., 2017). The monolayer NbSe2 is composed of an atomic Nb sandwiched by two Se, forming a trigonal prismatic crystal structure. Its remarkable and unique properties (Xi et al., 2015; Forro, Shan, & Mak, 2015; Bischoff et al., 2017; Ugeda et.al., 2016) lead to many theoretical and computational studies about the nature of its interaction with foreign atoms. Figure 1 shows the structure of monolayer *2H*-$NbSe_2$ (Silva-Guillén, Ordejón, Guinea, & Canadell, 2016) (we will call it $NbSe_2$), where *H* refers to hexagonal.

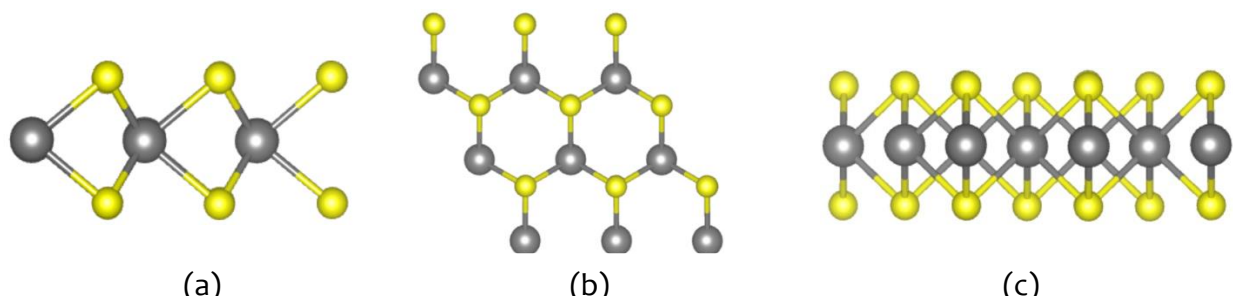

(a)                     (b)                    (c)

*Figure 1.* (a) Front, (b) top, and (c) side views of monolayer *2H*-$NbSe_2$.

Impurities in NbSe$_2$ are important to be studied because of the roles they play to modify some of its properties (Lawlor, 2018). Nguyen et al. (2017) investigated the introduction of C, N, and O impurities to defected monolayer two-dimensional NbSe$_2$ to better understand how it reacts with these atoms. Calculations reveal that these elemental impurities can stabilize the Se divacancies. A study conducted by Papageorgopoulos (1978) reveals that Cs and O$_2$ coadsorption with the NbSe$_2$ makes the resulting system very active in O$_2$ studies. However, despite these studies about O and O$_2$ interaction with NbSe$_2$, no study has been conducted to generally investigate the electronic properties of the resulting complex system.

In this work, we considered the investigation of the adsorption of O and O$_2$ to 2*H*-NbSe$_2$ hexagonal and calculate the charge transfer and electronic structures of the resulting complexes.

**METHODOLOGY**

All calculations were done using the First principles calculations (Ding et al., 2011) in Quantum Espresso 6.7. The exchange correlation was described by the Perdew-Burke-Ernzerhof (PBE) within the Generalized Gradient Approximation (GGA) (Wills, 2015). The Van der Waals interactions were considered by the DFT-D3 correction introduced in Moellmann and Grimme (2014). After convergence testing, we set the energy cut-off to 42 Ry.

The structures were optimized with a convergence criterion of $10^{-08}$ au. We allow them to fully relax until the individual ionic forces are less than $1.0 \times 10^{-03}$ eV/au. We used 3 x 3 x 1 supercell to mimic the structures of pristine, O, and O$_2$ adsorbed monolayer NbSe$_2$. To eliminate the interaction of neighboring monolayers of NbSe$_2$, we introduced a 13 Å vacuum slab towards the z-axis. We made sure that the O and O$_2$ are properly adsorbed by testing different sites using interstitial doping, and by varying their distance from the monolayer from 0 to 3 Å with respect to the z-axis above the hollow site and top sites of Nb and Se.

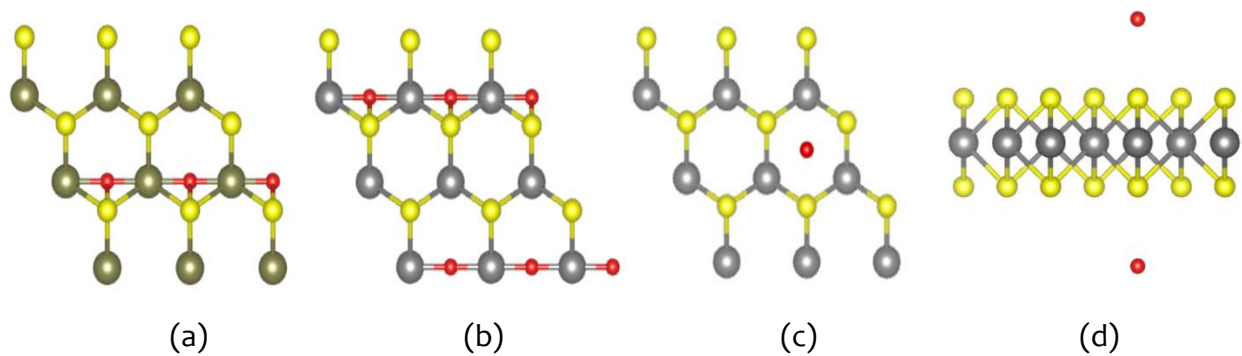

(a)  (b)  (c)  (d)

*Figure 2.* Some configurations considered for O-NbSe$_2$ complex showing (a) 3 interstitial O doping, (b) 6 interstitial O doping, (c) O adsorbed at the hollow site, and (d) Two O atoms adsorbed at the hollow site in the +/- z-direction.

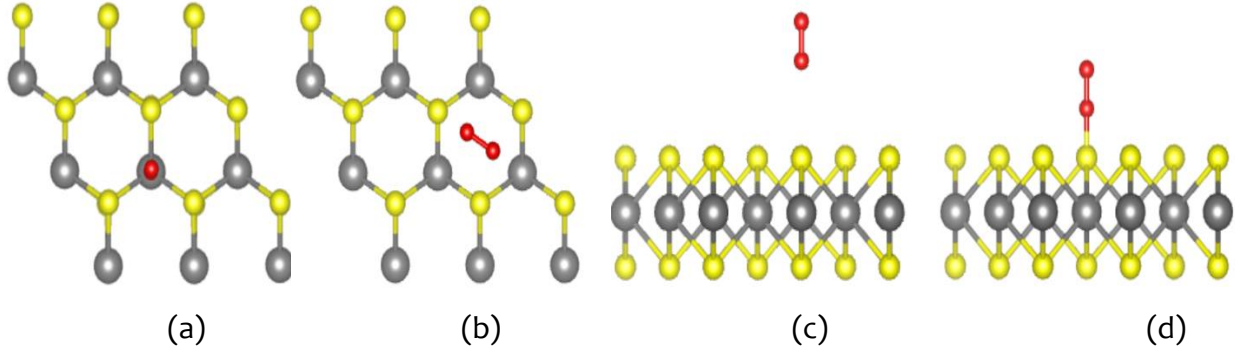

|  (a)  |  (b)  |  (c)  |  (d)  |

*Figure 3.* Some configurations considered for the $O_2$-$NbSe_2$ complex showing (a) $O_2$ at the top site of Nb, (b) $O_2$ at the hollow site, (c) $O_2$ at the top site of Se, and (d) $O_2$ partially attached to Se.

To obtain the stability of our structures, we calculated the binding energies using

$$E_b(pristine) = E_{NbSe_2} - E_{Nb} - 2E_{Se} \qquad \text{Equation 1}$$

for the pristine $NbSe_2$ where $E_{Nb}$ and $E_{Se}$ are the energies of the isolated Nb and Se respectively. For the O adsorbed $NbSe_2$,

$$E_b(O\ adsorbed) = \frac{E_{O_nNbSe_2} - E_{NbSe_2} - nE_O}{n} \qquad \text{Equation 2}$$

where $E_{ONbSe_2}$ and $E_{NbSe_2}$ are the energies of the O adsorbed and pristine $NbSe_2$, $E_O$ is the energy of isolated O atom and n is the number of H. For the $O_2$ adsorbed $NbSe_2$,

$$E_b(O_2\ adsorbed) = E_{O_2NbSe_2} - E_{NbSe_2} - E_{O_2} \qquad \text{Equation 3}$$

where $E_{O_2NbSe_2}$ and $E_{NbSe_2}$ are the energies of the $O_2$ adsorbed and pristine $NbSe_2$, and $E_{O_2}$ is the energy of the isolated $O_2$.

To find the magnitude and direction of the charge transfer, we calculate the charge density difference (CDD) (Tozini, Forti, Gargano, Alonso, & Rubiolo, 2015) and Bader charge analysis (Tang, Chill, & Henkelman, 1970).

## RESULTS

After full structural optimization of the considered complexes, we found that interstitial doping of O into the $NbSe_2$ could break the symmetry and the formation of the Nb and Se leading to the scattering of the atoms throughout the simulation box. Thus, in this study, we only considered the adsorption sites giving stable geometries.

After the calculation of the binding energies, we then proceeded to the investigation of the charge redistribution of the complexes. The following figures depict the charge density difference (CDD) of the different complexes where cyan and yellow isosurfaces indicate charge depletion and accumulation respectively.

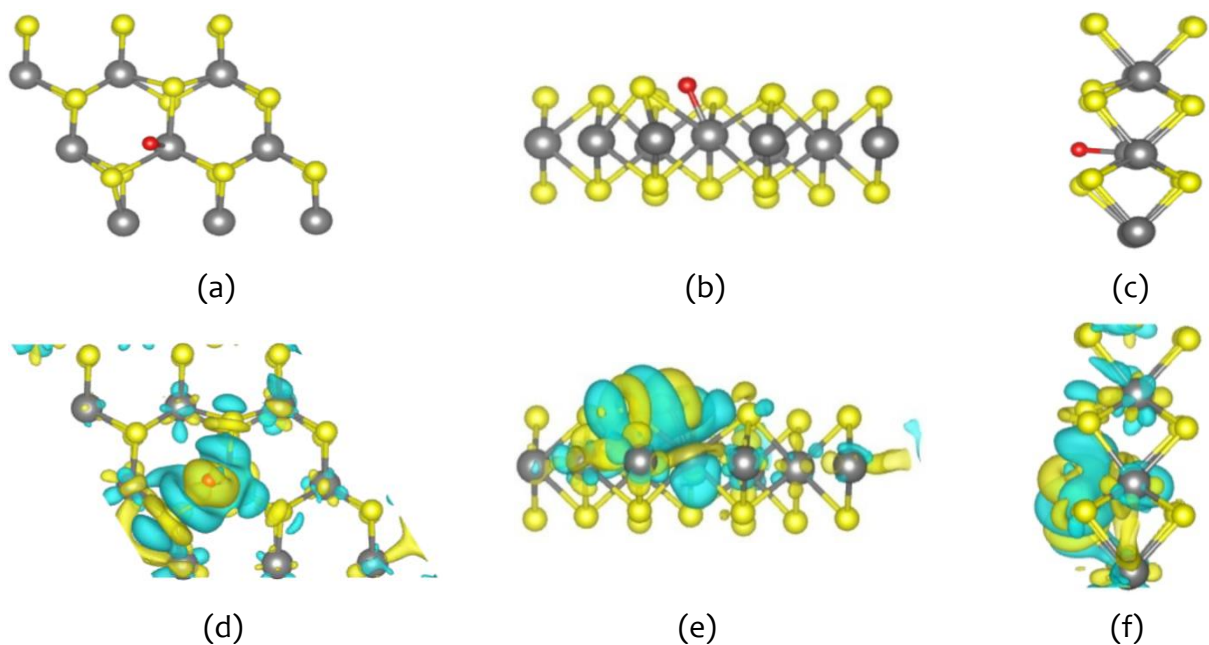

Figure 4. CDD of O-NbSe$_2$ complex showing the optimized (a,d) top, (b,e) side, and (c,f) front views of O adsorbed at the top site of Nb.

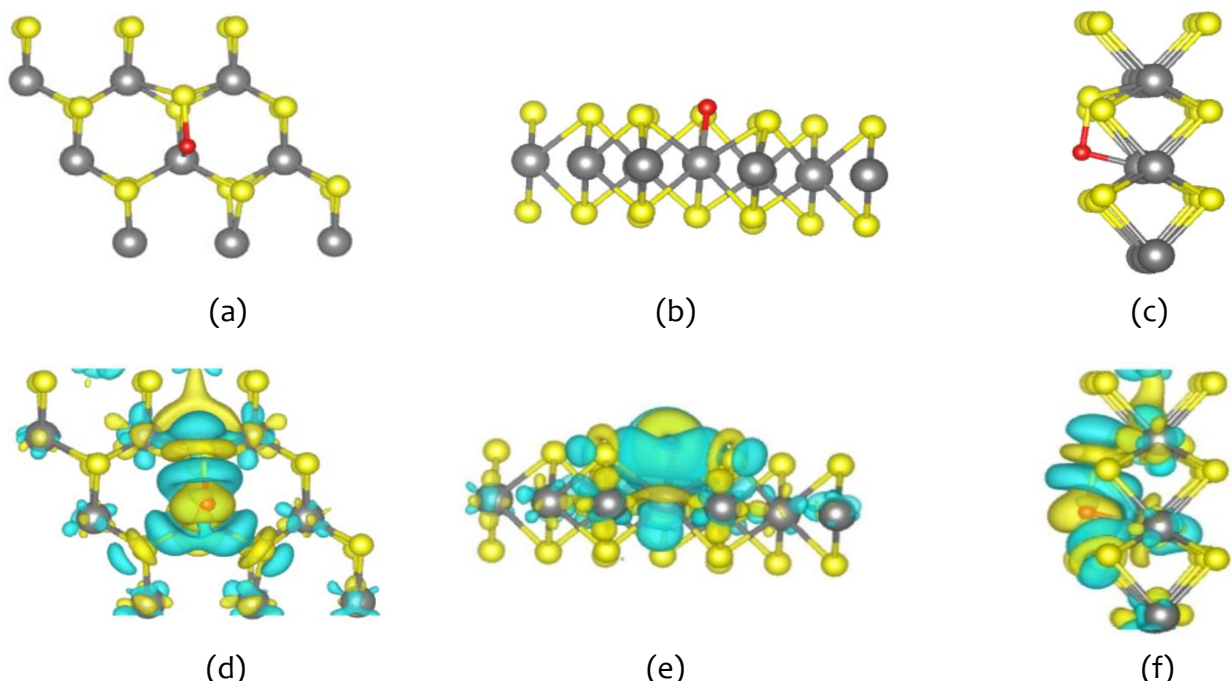

Figure 5. CDD of O-NbSe$_2$ complex showing the optimized (a,d) top, (b,e) side, and (c,f) front views of O adsorbed at the top site of Nb/Se.

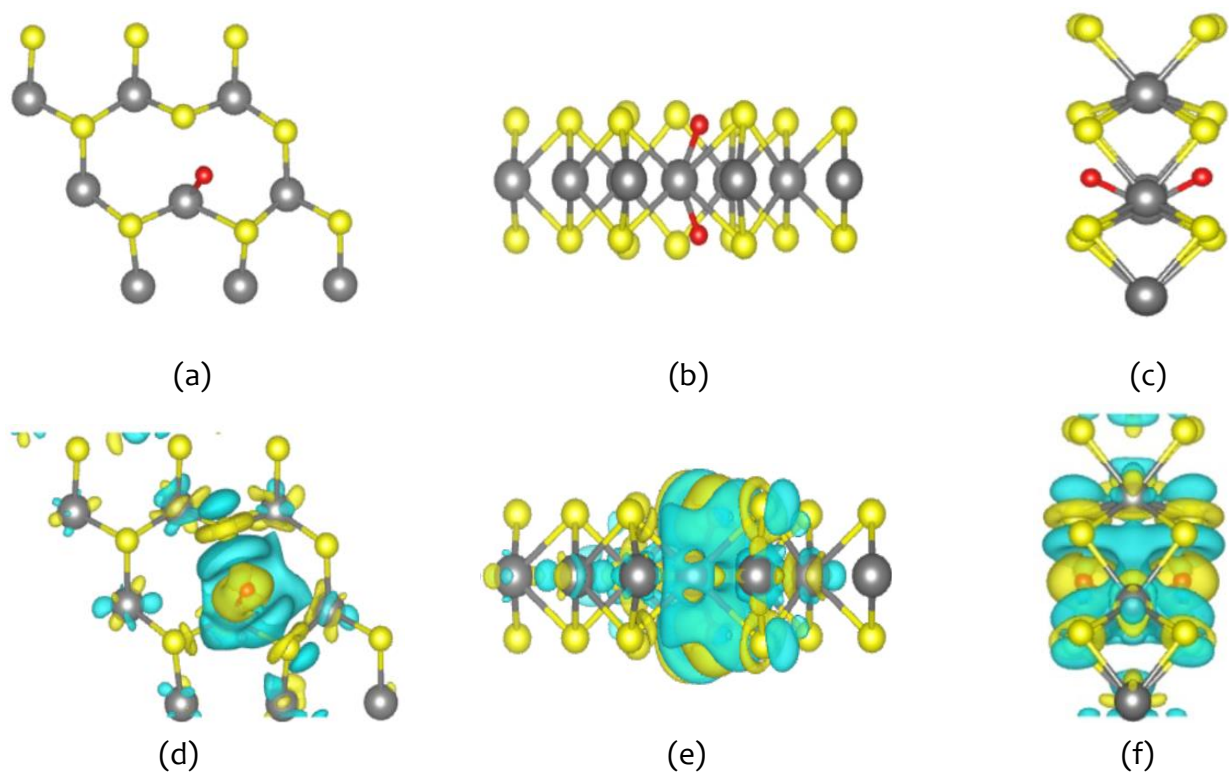

Figure 6. CDD of O-NbSe$_2$ complex showing the optimized (a,d) top, (b,e) side, and (c,f) front views of O adsorbed at the top site of Nb at +/- z-axis.

Using the Bader charge analysis, we were able to find the magnitude and direction of the charge transfer. Table 1 is the tabulated charge transfer for the different O-NbSe$_2$ complexes.

Table 1. Charge transfer of the different optimized configurations for O adsorption on monolayer NbSe$_2$ using Bader charge analysis.

| Configuration | Charge transfer |
| --- | --- |
| 1. Nb top site (standing) | 0.9193 $e$ |
| 2. Nb/Se top site (standing) | 0.9777 $e$ |
| 3. Top Nb (+/-z) | 1.8914 $e$ |

After the investigation of O-NbSe$_2$ complexes, we then considered the adsorption of O$_2$ to NbSe$_2$. We calculated first the binding energies for different adsorption sites to find out the most stable geometry. Figure 7 shows the plot of the binding energies while Table 2 shows the calculated values.

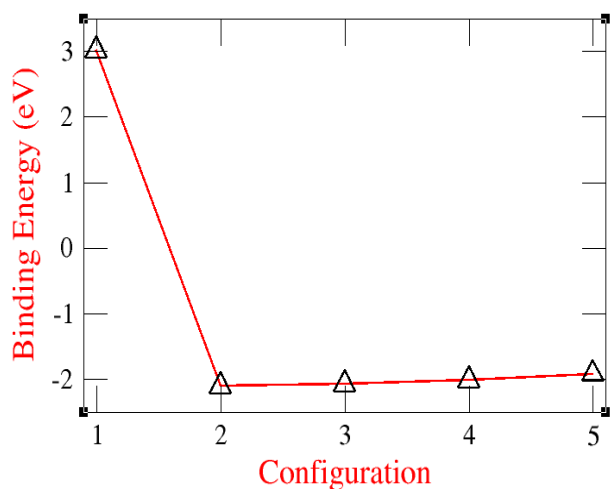

Figure 7. A plot of the binding energies of the different optimized configurations for $O_2$ adsorption on monolayer NbSe2.

Table 2. Binding energies of the different optimized configurations for $O_2$ adsorption on monolayer NbSe2.

| Configuration | Binding energy |
|---|---|
| 1. Hollow site (horizontal) | 3.01 eV |
| 2. Se top site (standing) | -2.11 eV |
| 3. Nb top site (standing) | -2.08 eV |
| 4. Nb (horizontal) | -2.02 eV |
| 5. Hollow site (standing) | -1.93 eV |

Based on the calculated binding energies, the most stable configuration is $O_2$ (standing) that is initially attached at the top site of Se. The following figures show the CDD of the $O_2$-NbSe$_2$ complexes. Here, we did not include the CDD of configuration 1 since the interaction is an endothermic process, indicating an unstable geometry.

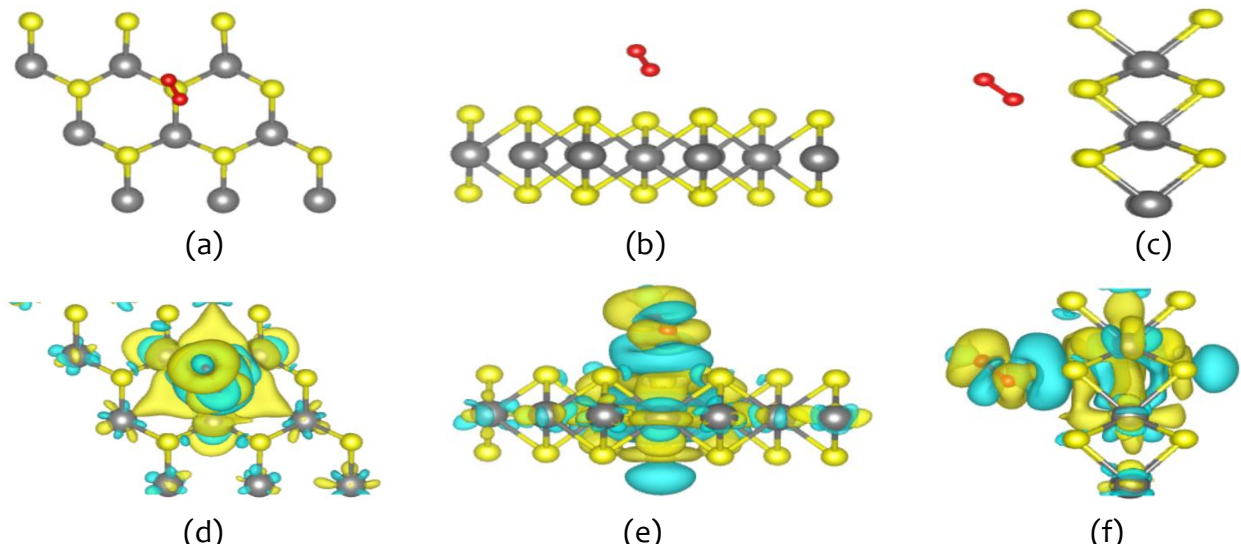

Figure 8. CDD of $O_2$-NbSe$_2$ system showing the optimized (a,d) top, (b,e) side, and (c,f) front views of $O_2$ adsorbed at the top site of Se.

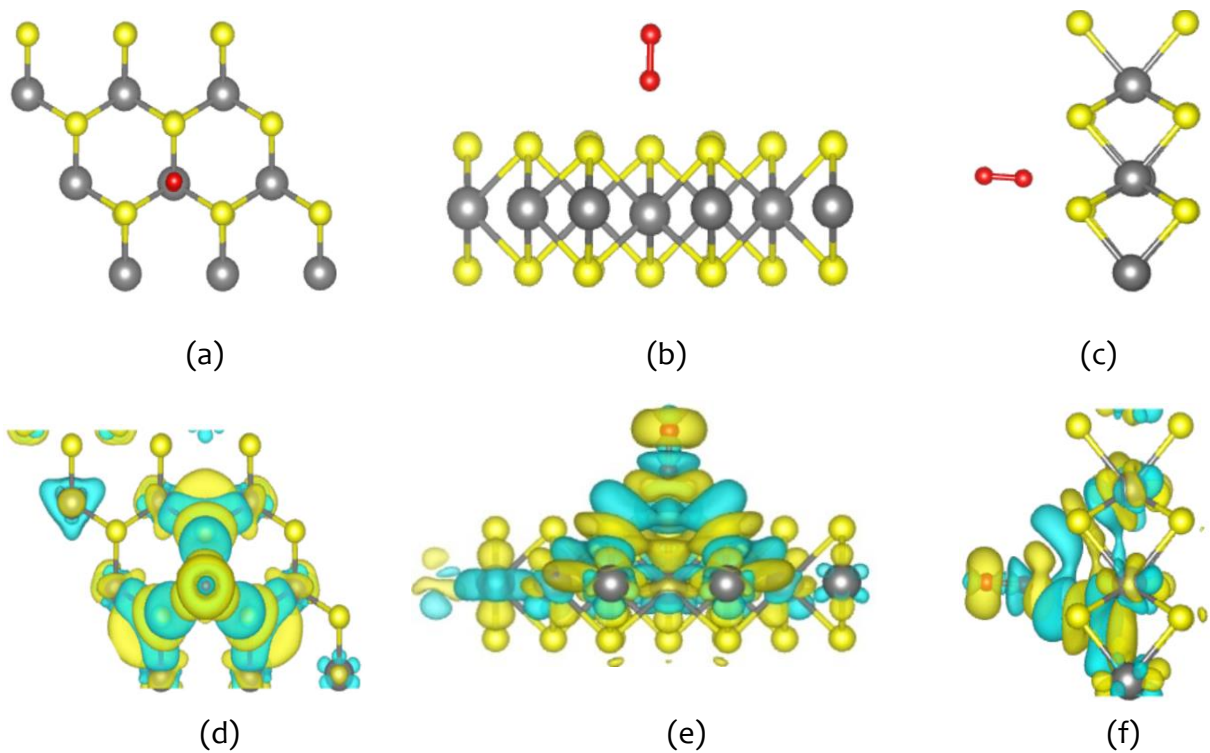

*Figure 9.* CDD of NbSe$_2$/O$_2$ system showing the optimized (a,d) top, (b,e) side, and (c,f) front views of O$_2$ adsorbed at the top site of Nb (standing).

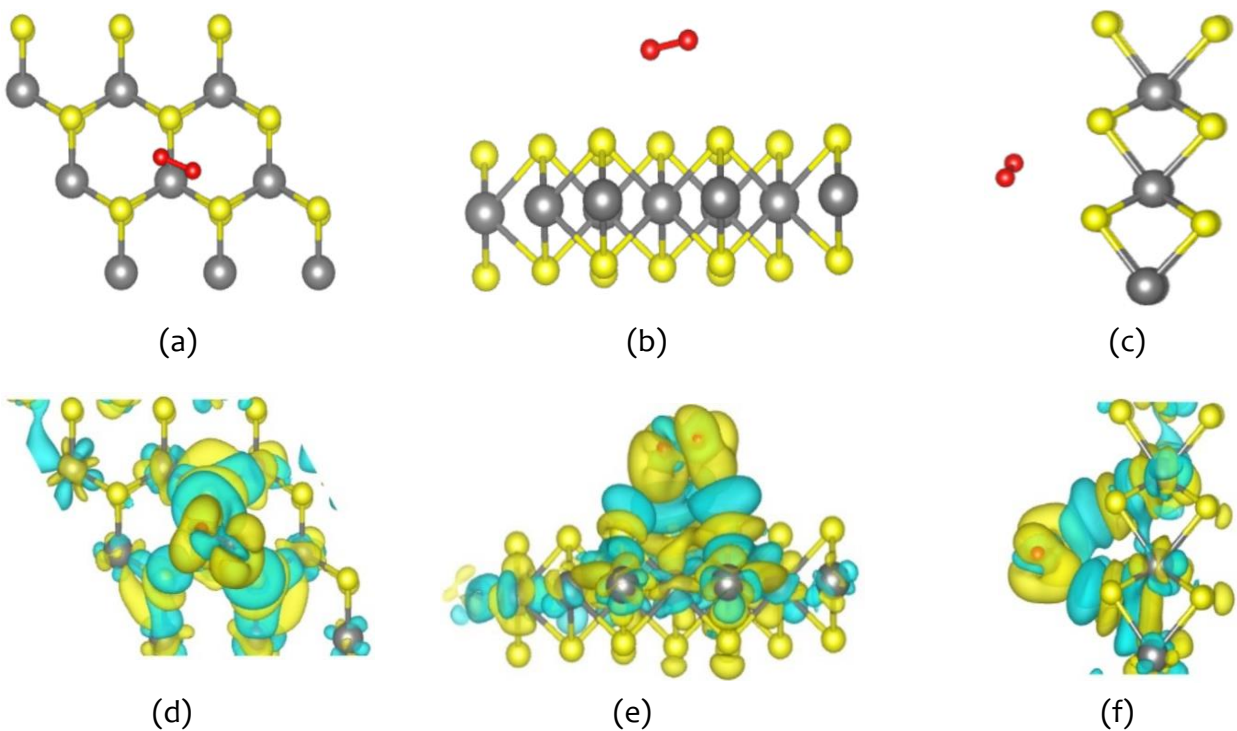

*Figure 10.* CDD of NbSe$_2$/O$_2$ system showing the optimized (a,d) top, (b,e) side, and (c,f) front views of O$_2$ adsorbed at the top site of Nb horizontally.

969

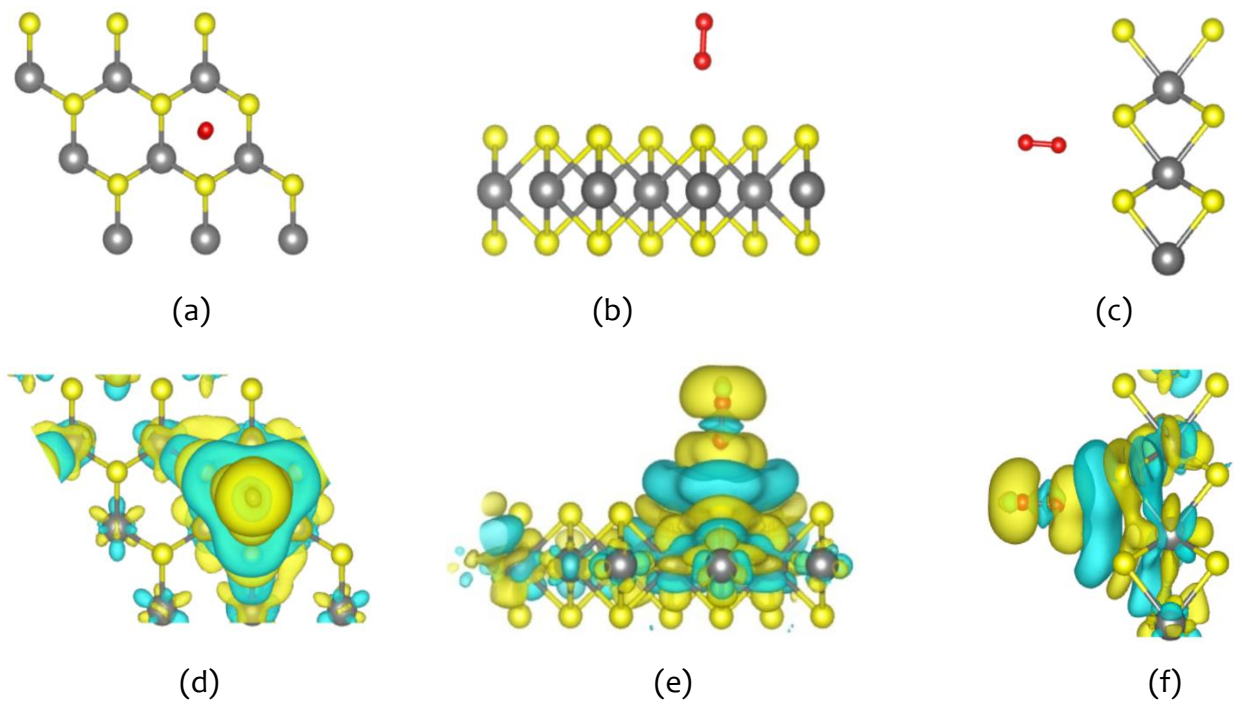

*Figure 11.* CDD of NbSe$_2$/O$_2$ system showing the optimized (a,d) top, (b,e) side, and (c,f) front views of O$_2$ adsorbed at the hollow site (standing).

Table 3 is the shows the magnitude of the charge transfer in O$_2$-NbSe$_2$ complexes as calculated by Bader charge analysis.

Table 3. Charge transfer of the different optimized configurations for O$_2$ adsorption on monolayer NbSe$_2$ using Bader charge analysis.

| Configuration | Charge transfer |
| --- | --- |
| 1. Se top site (standing) | 0.2171 e |
| 2. Nb top site (standing) | 0.1408 e |
| 3. Nb (horizontal) | 0.0883 e |
| 4. Hollow site (standing) | 0.0651 e |

We then analyzed the electronic structures of the O-NbSe$_2$ and O$_2$-NbSe$_2$ complexes with the most stable geometries and compare them to the electronic structures of pristine NbSe$_2$.

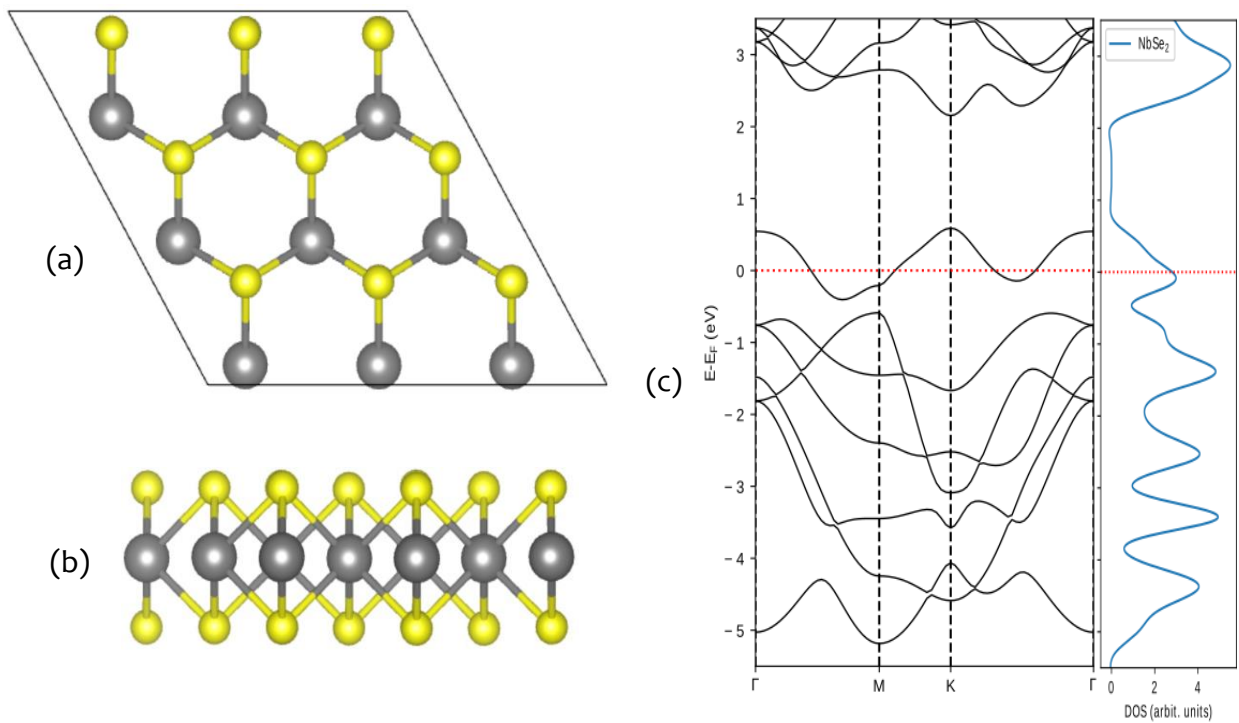

*Figure 12.* (a) Top and (b) side views of isolated pristine 2*H*-NbSe$_2$ and its (c) electronic structures.

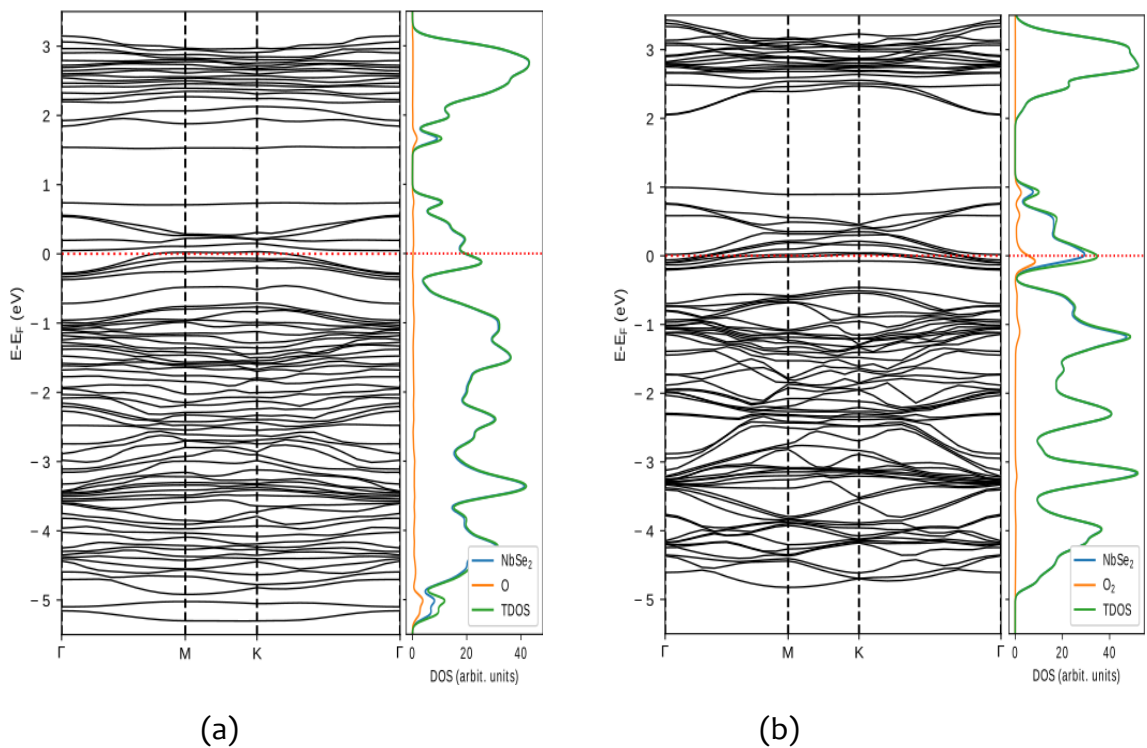

*Figure 13.* Electronic structures of optimized O-NbSe$_2$ (Nb top site) and (b) O$_2$-NbSe$_2$ (Se top site)

# DISCUSSION

In this work, we considered different adsorption sites for both O and $O_2$ to find which of the configurations has the most stable geometry. Accordingly, calculations reveal that interstitial doping is not practical for the system as it breaks the symmetry and formation of the atoms in the complexes.

For O-$NbSe_2$ complexes, we observed that the O atom tends to stabilize most at the top site of the Nb in the monolayer with an Nb-O bond length of 1.93 angstrom which is close with the previous works (Hardcastle & Wachs, 1991; West & Davies, 2011). Calculations reveal that the binding energy is -6.37 eV which is very close to the findings of Tafen & Gao (2013).

To predict the charge redistribution of O-$NbSe_2$ complexes, we calculated the Charge Density Difference (CDD). We observed that there is a significant charge transfer between the O and the monolayer. As a general feature, in Figure 4, large charge depletion (cyan) is observed from the atoms near the adsorption site, especially the Nb in the Nb-O bond. Furthermore, the O atom acted as an oxidizing agent (Liu et.al., 2020; Gholizadeh, & Yu, 2015) as revealed by the charge accumulation (yellow), consistent with established works. Calculations reveal that the net charge transfer is towards the O with a magnitude of 0.9193$e$. The same trend happens for Figure 5 with a charge transfer of 0.9777$e$ and Figure 6 with a charge transfer of 1.8914$e$. It is noted that the charge transfer when we adsorbed O atoms in the +/- z-axis is almost doubled indicating the proportionality of charge accumulation to the number of O atoms.

We then considered the adsorption of $O_2$ into the $NbSe_2$. Figure 7 shows the plot of the binding energies for different adsorption sites in the complexes. As shown, in Table 2, when $O_2$ is adsorbed horizontally at the hollow site of the monolayer, the interaction becomes endothermic, hinting at an unstable geometry with a binding energy of 3.01 eV. Among all the tested configurations, the $O_2$ tends to stabilize most at the top site of Se with a binding energy of -2.11 eV which is close to the Se-O binding energy calculated in the published papers (Ji, Xia, & Xu, 2016).

We also calculated the CDD of the $O_2$-$NbSe_2$ complexes. We can see in Figure 8 that there is a spontaneous charge transfer between $O_2$ and the monolayer. Just like the atomic oxygen, $O_2$ acted as a charge receiver with a magnitude of 0.2171$e$. Figures 9,10 and 11 also show the same direction for the charge transfer. Unlike the O-$NbSe_2$, $O_2$-$NbSe_2$ complexes have smaller magnitudes of charge transfer as seen in Table 3. This could be attributed to the nature of charge redistribution in the structures. As we know, the electronic charges in the individual O atoms in the molecular oxygen tend to redistribute as well as $O_2$-$NbSe_2$ complexes undergo charge transfer. As a result, the net charge accumulated may have been reduced. In all circumstances, for both the atomic and molecular oxygen adsorption to $NbSe_2$, the oxygen tends to receive electrons that may be attributed to its high electron affinity and electronegativity as compared to Nb and Se.

The electronic structures of the most stable complexes are also analyzed. Figure 12b shows the band structure and DOS of pristine $2H$-$NbSe_2$. As we can see, the Fermi level touches the conduction band with a nonzero DOS, indicating the metallic behavior of the monolayer (Silva-Guillén et al., 2016). In Figure 13a, the metallicity is preserved after O adsorption which is depicted by the band structure and relatively high density of states (DOS) near the Fermi energy level. We can observe that the attachment is not caused by hybridization of orbitals, consistent with the small magnitude of charge transfer, hinting at the possibility of physical binding (Koumpouras & Larsson, 2020) between O and $NbSe_2$. Further, at nearly -5 eV in the valence region, O caused a peak increase of the DOS suggesting a rearrangement in the valence band after O adsorption. Figure 13b depicts the electronic structures of the $O_2$-$NbSe_2$ complex. As observed, the metallic behavior of $NbSe_2$ is also retained. Moreover, we can see a peak increase near the Fermi level in the DOS caused by the $O_2$. Correspondingly, these findings are very useful for the determination of the effects of adsorbing an electron acceptor on $NbSe_2$ and other metallic nanostructures.

## CONCLUSIONS AND RECOMMENDATIONS

We have seen that oxygen adsorption to $NbSe_2$ preserves the metallicity of the monolayer $NbSe_2$. Calculations of the magnitude and direction of the charge transfer in the complexes confirm that atomic and molecular oxygen acts as oxidizing agents when interacting with metals.

In all circumstances, our study provided a theoretical prediction about the effect of oxygen on the stability and electronic properties of $NbSe_2$. The calculated binding energies show the firmness of the structures which could lead the way to tailor the electronic behavior of $NbSe_2$ to understand more the nature of its possible applications in nanotechnology and other nanoelectronics-related devices. To have a deeper understanding of the interaction of these materials, further investigations should be conducted like calculation the electron localization function to confirm the type of interaction and calculating the optical and vibrational properties.

## IMPLICATIONS

This work is done to have an insight into the electronic properties of oxygen-$NbSe_2$ complexes. As known, the interaction of atoms and molecules to pristine materials is very important to have wider insights into its properties and potential applications. Thus, in this study, we adsorbed atomic and molecular oxygen atoms to monolayer $NbSe_2$ to provide theoretical predictions about the electronic properties of the resulting complexes. Accordingly, the results are consistent with the published work and articles, hinting that our calculations are accurate.

## ACKNOWLEDGEMENT

The authors would like to thank the Department of Physics in MSU-Marawi for the moral support and encouragement.